\documentclass[useAMS,usenatbib]{mn2e}
\usepackage{natbib}
\usepackage{epsf}
\usepackage{epsfig}
\usepackage{times}
\usepackage{ulem}
\usepackage{mathrsfs}
\usepackage{color}
\usepackage{aas_macros}
\usepackage[utf8x]{inputenc}
\usepackage{graphicx}
\usepackage{amssymb}
\usepackage{ifthen}
\usepackage{caption}
\usepackage{subcaption}
\usepackage{amsmath}
\def\swift{{\it{Swift}}~}
\def\swiftns{{\it{Swift}}}
\def\t90{$T_{90}$}

\newcommand{\gae}{\lower 2pt \hbox{$\, \buildrel {\scriptstyle >}\over {\scriptstyle \sim}\,$}}
\title{The rate, luminosity function and  time delay of non-Collapsar short GRBs}
\author[D. Wanderman \& T. Piran]{{David Wanderman$^1$ and Tsvi Piran$^1$}\\ $^1${Racah Institute of Physics, The Hebrew University, Jerusalem 91904, Israel}}
\begin{document}
\maketitle
\begin{abstract}
We estimate the rate and the luminosity function of short (hard) Gamma-Ray  Bursts (sGRBs) that are non-Collapsars, using the peak fluxes and redshifts of BATSE, Swift and Fermi GRBs. 
Following \cite{Bromberg2013} we select  a sub-sample of \swift bursts which are most likely non-Collapsars. We find that these  sGRBs are  delayed relative  to the global star formation rate (SFR) with a typical delay time of a $3-4 $~Gyr (depending on the SFR model). 
 {  However, if two or three sGRB at high redshifts have been missed because of selection effects, a distribution of delay times of $\propto 1/t $ would be also compatible. } The current event rate of these non-Collapsar sGRBs with $L_{iso}>5\times 10^{49} erg/s$ is $4.1_{-1.9}^{+2.3}Gpc^{-3}yr^{-1}$.
The rate was significantly  larger  around  $z \sim 1$ and it declines since that time.
The luminosity function we find is a broken power law with a break  at $2.0_{-0.4}^{+1.4} \times 10^{52}$~erg/s and power-law indices $0.95_{-0.1 2}^{+0.12}$ and $2.0_{-0.8}^{+1.0}$.
When considering the whole \swift sGRB sample we find that it is  composed  of two populations: One group ($\approx 60\%-80\%$ of \swift sGRBs) with the above rate  and time delay and a second group ($\approx 20\%-40\%$ of \swift sGRBs) of potential ``impostors" that follow the SFR with no delay. These two populations are in very good agreement with the division of sGRBs to  non-Collapsars and Collapsars suggested recently by \cite{Bromberg2013}.
If non-Collapsar sGRBs arise from neutron star merger this  rate suggest a detection rate of  3-100 yr$^{-1}$ by a future gravitational wave detectors (e.g. Advanced Ligo/Virgo with detection horizon on 300 Mpc), and a co-detection with Fermi (\swift) rate of  0.1-1 yr$^{-1}$ (0.02-0.14 yr$^{-1}$). 
We estimate that about $4 \times 10^5 (f_b^{-1}/30)$ mergers took place in the Milky Way.  If  $0.025 m_\odot$ were ejected in each event this would have been sufficient to produce all the heavy r-process material in the Galaxy.
\end{abstract}

\section{Introduction}
GRB are  divided into two distinct groups: long bursts with $T_{90}>2 $~sec\footnote{  T90 is the duration in which the central 90\% of the gamma-ray signal is detected.} and short ones with $T_{90}<2 $~sec \citep{Kouveliotou1993c}. The short bursts are also typically harder than  the long ones \citep[][]{Dezalay1996a,Kouveliotou1996m}.    Using the $\langle V/V_{max}\rangle$ test\footnote{$<$V/Vmax$>$ is the ratio of the volume enclosed by a burst (a sphere with radius equals to the burst's distance), and the volume in which such a burst could be detected. Different $<$V/Vmax$>$ values for the samples implies that the two samples have a different spatial distribution.}, \cite{Cohen1995k} have shown  that the two populations are distributed differently in redshift space with the short GRBs being typically nearer to us, but still within cosmological distances. 
{  Afterglow detection in the late 90ies confirmed the cosmological origin of long GRBs and together with the discovery of SNe associated with the GRBs led to their identification as Collapsars.}
The origin of sGRBs remained obscure until  2005 when the { HETE-2} and {\it Swift} satellites detected   the first 
 sGRB  afterglows from GRBs 050509b, 050709 and 050724 (Bloom et al. 2006; D. B. Fox et al. 2006; Gehrels et al. 2005; Castro-Tirado et al. 2005b; Prochaska et al. 2005; Fox et al. 2005; Hjorth et al. 2005a, 2005b; Covino et al. 2006; Berger et al. 2005).  This has led to the identification  of the host galaxies. The difference between those hosts and the hosts   of long bursts  
 provided a strong indication that sGRBs result from a different type of physical progenitors{   \citep[e.g.][]{Berger2014, Fong2013a, Nakar2006}.}  When more sGRB redshifts have been accumulated the 
 relatively low redshifts $\bar z \sim 0.5$ confirmed the earlier expectations, that were based just on 
 the peak flux distribution,  that the short bursts are nearer on average than the  long ones (whose average redshift satisfies $\bar z >2$). 

{  Recently \cite{Bromberg2013} have shown that the \swift sample of short ($T_{90}<2 $~sec) bursts is contaminated by a significant  fraction of short duration Collapsars \citep[See however][who seem to arrive to a different conclusion based on host morphologies]{Fong2013a}.} They have argued that due to its soft  energy window \swift is less sensitive to the harder sGRBs and hence  the division at $T_{90}=2 $~sec found for BATSE bursts is inadequate for the \swift  sample.  They  have estimated the probability that a given \swift short duration burst with a known duration and hardness is a non-Collapsar or that it is a short Collapsar.  \cite{Bromberg2013} divide the bursts to subgroups according to  their spectral index and estimate this probability  as a function of the burst's duration for each group separately. 
In this work we use  this method   to build a  sample of non-Collapsars sGRBs in a way which minimizes the sample contamination by Collapsars  and allows us to put a handle on the uncertainty associated with the sample contamination. 
Using this sample we explore the rate of non-Collapsar sGRBs and their luminosity function. 
We will refers in the rest of the text to this sample, so when we discuss sGRBs we actually mean non-Collapsar sGRBs, unless specified otherwise. 
We also explore the implication of using the full (contaminated sample) and we find that indeed the division, as suggested by  \cite{Bromberg2013}, is well justified.  

The luminosity function and formation rate of sGRBs are fundamental to understanding the nature of these objects.  
It has been suggested \citep{eichler1989} that sGRB originate from neutron star-neutron star (ns$^2$) or neutron star-black hole (nsbh) mergers. In this case we expect a delay between the SFR and the  merger rate \citep{Piran1992b,Ando2004bi} due to the time it takes for the binary orbits to spiral in. As such  
information about the rate could reveal the magnitude of this delay and provide a clue on the physics of the progenitor system, the relevant information goes beyond the physics of GRBs. 
The mergers end with a chirping burst of gravitational waves (GW) and gravitational detectors  whose prime goal is  to detect those signals are upgraded (LIGO, Virgo) or constructed (KAGRA) now. A major unknown factor in these observations is the expected rate of mergers within the detection horizon.  
The local sGRB rate is important for  estimating  of the detection rate of GW telescopes.  
Mergers are also a likely source of heavy r-process material \citep{Lattimer74,eichler1989}. A crucial question is whether they can provide all heavy r-process material or is there another significant heavy r-process source.  The major factor in answering this question is the  overall rate of sGRBs. However, the variation of this rate with time is also important as some r-process material was produced very early on and an interesting question is whether mergers (that resulted also in GRBs)  could have produced this material. {  Another important clue might be the recent discovery of macronova-like signature in GRB130603B \citep{tanvir2013j,Berger+13}. }

There is a large sample of sGRBs detected by  BATSE and by Fermi/GBM. However  the redshifts of these  bursts is unknown. The observed
peak flux   distribution of these bursts is a convolution of  two unknown functions, the rate and the luminosity function and  it is impossible to determine both functions without additional information.
For example, \cite{Cohen1995k} have shown that the observed BATSE flux distribution can be fitted with very different luminosity functions depending on the choice of the GRB rate and vice versa. 
Furthermore, \cite{Nakar2006} has shown that for a single power law the peak-flux distribution will also be a power law and it will not depend on the formation rate of the GRBs. Since the BATSE peak-flux data is indeed consistent with a single power law luminosity function we cannot use it to  constrain the formation rate.

We resolve this problem  by combining the small \swift \citep{Gehrels2004} sGRB  redshift sample with the large sGRB samples of BATSE and Fermi. We carry out a combined likelihood analysis using the BATSE and the   Fermi/GBM peak flux distributions and the \swift sGRB redshift sample. 
 The combined sample  enables  us to constrain both  the rate of sGRB and  the luminosity function.
In \S \ref{sec:samples} we define the  samples  and discuss possible  selection effects. We define in \S \ref{sec:method}  models for the luminosity function and the rate, considering a time delay with respect to the SFR. We also build the likelihood function and describe the method to determine the most probable parameters for the rate and the luminosity function and their associated uncertainty range. In \S \ref{sec:results} we present the results of the analysis for the different models and finally in \S \ref{sec:discussion} we discuss the results and their implications on the progenitors of sGRBs and on the detection rate of gravitational waves. We also discuss the consistency of our results with the predicted fraction of Collapsars and non-Collapsars suggested by \cite{Bromberg2013}.

\section{The samples}
\label{sec:samples}
{  We jointly analyze here observations from three major GRB targeted missions: BATSE, \swift and Fermi.
We stress that there is no single standard way to jointly analyze the observations of the different detectors having different energy bands and different triggering scheme. We must make certain assumptions to allow such joint analysis: first we define a threshold for the duration for each detector’s sample in a consistent way. Obviously setting the same value of \t90 threshold for all detectors does not serve this goal since the \t90 are defined differently on the detectors (\t90 is defined for the energy bands [50-300]keV, [15-150]keV and [50-300]keV for BATSE, \swift and Fermi respectively). A more consistent way would be taking a uniform threshold value of $f_{NC}$ for all detectors i.e. taking only bursts which are more likely to be a non-collapsars (e.g. $f_{NC} >$ 0.6, as we used for \swift). 	
}

Our samples consist of short  BATSE and Fermi bursts with a measured peak-flux and of \swift short bursts with redshift\footnote{We do not include the \swift peak-flux (without redshifts) sample as it is much smaller ($\sim 40$ bursts) than the corresponding BATSE and Fermi samples.}. Following \cite{Bromberg2013} we estimate the level of contamination of the sample by Collapsars events. For the BATSE and Fermi we adopt the traditional duration cutoff $T_{90}<2 $~sec as it yields a low contamination of Collapsars in the sample at the  $10\%-15\%$ level. For the \swift sample a duration cutoff of $T_{90}<2 $~sec would yield a contamination fraction of $32\%$, hence we adopt the more sophisticated selection method using both the duration and the spectral index of the bursts. We  include bursts for which the  probability of being non-Collapsar  is $>0.6$. This results in a sample of 12 bursts with an estimated contamination fraction of only $7\%$ i.e. only one of these bursts is expected to be an impostor (see 
table \ref{tbl:sGRBs_sample} for the sample).
{  The contamination fractions for the different detectors are comparable (10\%, 15\%,7\%).}
{  The use of different contamination fractions has no impact on the major results in this paper.}

There are number of selection effects concerning the peak-flux and the redshift samples \citep[see][for a review]{Coward2013}.
We minimize the biases that arise  from the detectors' sensitivity limit by selecting a high flux limit for BATSE and Fermi bursts.
The \swift redshift sample is small and hence we use the whole sample, therefore it is susceptible to the biases described in \cite{Coward2013}. 
We caution though that if the lack  of higher redshift ($z>1.2$) non-Collapsar sGRBs is a result of an observational bias    some of our conclusions concerning the delay between the sGRBs and the SFR (see  \S \ref{sec:discussion}.)  might be significantly weakened. {  In particular we discuss later a case in which  a few bursts at $z>1.2$ are artificially added to the sample and check its implications.}

\subsection{The BATSE and Fermi Flux samples}
Following \cite{Nakar2006} we extract from the current BATSE catalog\footnote{http://www.batse.msfc.nasa.gov/batse/grb/catalog/current/} all  short ($T_{90}<2 $~sec) bursts with a peak flux in the 64 ms timing window of $P_{64} > 1.5$ [ph/cm$^2$ sec], yielding a sample of 341 bursts. The duration $T_{90}=2 s$ corresponds to a non-Collapsar probability of 0.6, and the expected fraction of non-Collapsar in the entire sample   is 0.9.

We take the short ($T_{90}<2 $~sec) bursts with a peak flux in the 64 ms timing window (and in the [50keV - 300keV] energy band) of $P_{64} > 2.37$ [ph/cm$^2$ sec] from the Fermi GBM Burst Catalog\footnote{http://heasarc.gsfc.nasa.gov/W3Browse/fermi/fermigbrst.html} \citep[][]{Paciesas2012a,Paciesas2012b} until date 10-April-2013. 
This  yields a sample of 145 bursts. The flux threshold is  estimated by comparing the peak-flux distribution with that of BATSE. Consistent with the findings of \cite{Nava2011} the Fermi threshold is higher than the  BATSE threshold. $T_{90}=2 $~sec corresponds to a non-Collapsar probability  of 0.45, and the expected fraction of non-Collapsar in the entire sample  is 0.85.

\subsection{The Swift Redshift sample}
\label{sec:swift_sample}
The entire \swift sGRBs redshift sample consist of 20 short ($T_{90}<2 $~sec) \swift bursts which have a redshift determination. We  did not include bursts that have a short peak followed by an extended emission with overall $T_{90}>2 $~sec. The $T_{90}$ duration and the 1-sec peak-flux are taken from the \swift mission page\footnote{http://swift.gsfc.nasa.gov/docs/swift/archive/grb\_table/grb\_table.php}. 
The 64-ms peak-flux of the bursts was estimated by correcting the 1-sec peak-flux with the ratio of the 64-ms and 1024-ms peak-counts in the 64-ms binned light-curve provided in \swift Burst Ground-Analysis Information page\footnote{http://gcn.gsfc.nasa.gov/swift\_gnd\_ana.html} \citep[][]{Barthelmy2005a, Sakamoto2008, Sakamoto2011ax}.
The probability of a burst being a non-Collapsar is calculated following \cite{Bromberg2013}.
Table \ref{tbl:sGRBs_sample} lists, for the \swift bursts sample, the redshift, peak-flux, luminosity, duration, photon index and $f_{NC}$ - the probability for being a non-Collapsar.
From this sample we select a subgroups of 12 events with a high probability for being non-Collapsars ($f_{NC}>0.6$) to establish a ``genuine"  redshift sample on which we carry most of the analysis. We will return to the full sample in \S\ref{sec:met_twocomp} and establish that it is indeed more compatible with a bimodal distribution of  event rates.

The  Peak-flux threshold alone is a very rough estimate for  the \swift GRB detection threshold \citep[see e.g.][]{Lien2014c} as the actual detection as well as redshift determination depend on many factors. Nonetheless we have estimated the \swift detection threshold for sGRB with redshift measurement\footnote{There is only one burst with peak-flux below the $P_{lim}=2.5$ threshold, however since this threshold is only a rough estimate for the detection probability keeping this burst may compensate for the missed bursts above $P_{lim}=2.5$. We have repeated the analysis with and without this burst and found no significant change in the results. This limit was estimated as the peak-flux value for which the observed bursts number starts to significantly deviate from the relation $N(P)/dP \propto P^{-2}$ observed in BATSE \citep[][]{Nakar2006}.} 
roughly as $P_{lim}=2.5 ph/s/cm^2$.
We have repeated the analysis for other values of $P_{lim}$ and found that the luminosity function and the rate do not change significantly. However, the expected  luminosity distribution and redshift distribution are consistent with the observations only for $P_{lim}\simeq2.5$.
 
\begin{table}
\begin{tabular}{|l|l|l|l|l|l|l|l|}
\hline
GRB  & z & Ref. & $P_{64}$ & $L_{51}$ & $T_{90}$ & $P.L.$ & $f_{NC}$ \\
\hline \hline
050509B & 0.225 & 1 & 1.32 & 0.22 & 0.07 & 1.57 & 0.87 \\060502B & 0.287 & 2 & 3.24 & 0.89 & 0.13 & 0.98 & 0.99 \\060801  & 1.131 & 3 & 4.09 & 16.14 & 0.49 & 0.47 & 0.95 \\061201  & 0.111 & 4 & 10.46 & 0.42 & 0.76 & 0.81 & 0.92 \\061217  & 0.827 & 5 & 6.73 & 14.81 & 0.21 & 0.86 & 0.98 \\071227  & 0.383 & 6 & 4.99 & 2.44 & 1.80 & 0.99 & 0.71 \\080905A & 0.122 & 7 & 4.63 & 0.23 & 1.00 & 0.85 & 0.88 \\090510  & 0.903 & 8 & 46.75 & 121.4 & 0.30 & 0.98 & 0.97 \\100117A & 0.920 & 9 & 11.14 & 29.95 & 0.30 & 0.88 & 0.97 \\100206A & 0.407 & 10 & 11.91 & 6.56 & 0.12 & 0.63 & 0.99 \\101219A & 0.718 & 11 & 12.01 & 20.16 & 0.60 & 0.63 & 0.94 \\130603B & 0.356 & 12 & 47.30 & 20.01 & 0.18 & 0.82 & 0.99 \\\hline 
 Average & 0.532 &  & 13.71 & 19.44 & 0.50 & 0.87 & 0.93 \\

\hline \hline
050813$^a$  & 1.8$^a$ & 13 & 4.65 & 42.10$^a$ & 0.45 & 1.28 & 0.57 \\
051221A & 0.547 & 14 & 45.84 & 45.33 & 1.40 & 1.39 & 0.18 \\070429B & 0.904 & 15 & 7.20 & 18.73 & 0.47 & 1.72 & 0.32 \\070724A & 0.457 & 16 & 3.06 & 2.12 & 0.40 & 1.81 & 0.37 \\070809  & 0.219 & 17 & 2.62 & 0.42 & 1.30 & 1.69 & 0.09 \\090426  & 2.609 & 18 & 5.16 & 87.84 & 1.20 & 1.93 & 0.10 \\100724A & 1.288 & 19 & 4.14 & 20.72 & 1.40 & 1.92 & 0.08 \\131004A & 0.717 & 20 & 6.73 & 11.27 & 1.54 & 1.81 & 0.29 \\\hline 
 Average & 0.933 & & 9.93 & 24.29 & 1.02 & 1.69 & 0.25 \\\hline \hline 
 Total Avg. & 0.693 & & 12.20 & 21.38 & 0.71 & 1.20 & 0.66 \\
\hline
\end{tabular}
\caption{The \swift sGRBs redshift sample: \swift bursts with $T_{90} \lesssim 2 $~sec and a measured redshift. The top part includes the bursts with $f_{NC} > 0.6$ and the bottom part includes the bursts with $f_{NC} \leq 0.6$. $P_{64}$ is the 64-ms peak photon flux in $ph \ s^{-1} cm^{-2}$ in the \swift energy band [15keV - 150keV]; $L_{51}$ is the peak luminosity in $10^{51}erg/s$ (see text for the detailed definition);  $T_{90}$ is the observed burst duration in seconds;  $P.L.$ is the spectral photon power-law index in the \swift energy band [15keV - 150keV]; $f_{NC}$ is the probability that this bursts is a non-collapsar (genuine sGRBs)  given the burst's duration and its photon power-law index.  
(a) Two possible redshifts are suggested  for the burst 050813: 1.8 and 0.722. We list the first in the table with the corresponding luminosity and refer to the sample as full sample (A). The corresponding luminosity for this burst at z=0.722 is $7.89 \times 10^{51}$~erg/s and the sample with these values for that bursts is refered to as full sample (B).
Redshift references: (1) \citealp{Prochaska2005d,Gehrels2005j}; 
    (2) \citealp{Bloom2006ad}; 
  (3) \citealp{Cucchiara2006i}; 
 (4) \citealp{Berger2006aj,Berger2007ag}; 
  (5) \citealp{Berger2006bp}; 
(6) \citealp{D'Avanzo2007k,Berger2007ap}; 
  (7) \citealp{Rowlinson2010g};
 (8) \citealp{Rau2009l};  
(9) \citealp{Fong2011w}; 
 (10) \citealp{Cenko2010f};
 (11) \citealp{Chornock2011c};
 (12) \citealp{Thone2013,2013GCN..14745...1F,Sanchez-Ramirez2013a,Cucchiara2013n,2013GCN..14757...1X}; 
  (13) \citealp{Gehrels2005j,Berger2005cu,Foley2005f};
    (14) \citealp{Berger2005ar,Soderberg2006w}; 
   (15) \citealp{Perley2007r};
   (16) \citealp{Cucchiara2007d,Covino2007e}; 
 (17) \citealp{Perley2008d};
 (18) \citealp{Levesque2009j,Thoene2009d}; 
 (19) \citealp{Thoene2010};
 (20) \citealp{2013GCN..15307...1C}; \citealp{2013GCN..15310...1D}.
}
\label{tbl:sGRBs_sample}
\end{table}

\section{Methods}
\label{sec:method}
We construct models for the  luminosity function and the rate and estimate the likelihood for obtaining the observed data given these models. By maximizing the likelihood, we obtain the best fit parameters for the different models. We estimate, using the likelihood function and Monte Carlo simulations the errors in these parameters.  

\subsection{The models} 
\label{subsec:the_models}
The models are a functional forms for the luminosity function and for the rate. For the luminosity function we chose the standard broken power law used when considering GRBs:
 \begin{equation}
 \phi_0(L) =
\left\{
\begin{array}{ll}
(\frac{L}{L^*})^{-\alpha_L}   & L < L^* \ , \\
(\frac{L}{L^*})^{-\beta_L} & L > L^* \ .
\end{array}
\right. 
\end{equation}
This luminosity function is the \emph{logarithmic} luminosity function defined as the (un-normalized) fraction of bursts within a logarithmic  interval  $d \log( L)$. The ``linear`` luminosity function is related to the logarithmic one  by incrementing $\alpha_L$ and $\beta_L$ by 1. 
We designate this general  model as a broken power-law when $\alpha_L$, $\beta_L$ and L* are free parameters. 
We wish to examine the significance of the break in the luminosity function by also considering a model with a single power law luminosity function. By comparing the likelihood of the models, we show later, at the results section, that a broken power law is significantly preferred over a single power law for modeling the luminosity function.

We define the 64-ms isotropic-equivalent peak luminosity $L$ in the [1keV - 10MeV] energy band by assuming an average Band-function \citep{Band1993g} for all  bursts using the typical parameters of the Band function\footnote{  We have repeated the analysis another six times by separately varying each of the Band function parameters (i.e. $E_{peak}, \alpha_{BAND}, \beta_{BAND}$) by plus or minus its standard deviation. In all cases the effect on the final reaults is negligable: The log-normal time delay model is preffered over the power-law time delay model by almost the same likelihood ratio, and the most likely time delay parameter as well as its 1-$\sigma$ uncertainty range change by less than 0.2Gyr.
The only difference regards the 2-$\sigma$ and 3-$\sigma$ error range which are smaller for $\alpha_{BAND} = -0.1$ and larger for $\alpha_{BAND} = -0.9$, which marginally allows a zero time delay at the 3-$\sigma$ level.
} for Fermi/GBM \citep{Nava2011}:  $E_{peak}$~= 800 keV (in the source frame\footnote{
{  The source frame value of the typical Band function's $E_{peak} \sim$~800keV was estimated from the typical observed value in Fermi - $E_{peak}$~ = 490keV using the mean \swift redshift of our non-collapsars (\=z = 0.69). Obviously this is not fully self-consistent, and the \swift mean redshift was used only since the Fermi mean sGRB redshift is unknown. We can however use the results of our analysis to calculate the expected redshift value for the \cite{Nava2011} Fermi sample and examine a-posteriori the consistency of the $E_{peak}$ value adopted in our paper.
The value we have calculated for the Fermi non-collapsars is \=z = 0.27. 
Taking into account the contamination fraction of 15\% in Fermi sGRB sample, assuming for it the long GRB distributions, which have \=z $\sim$ 2, we estimate a value of \=z = 0.53 for the sample of Fermi/GBM used by \cite{Nava2011}.
Since this value is close to the \=z = 0.69 we had to begin with, we can assume that our method is self-consistent.
The effect of changing the spectrum $E_{peak}$, from 800 keV to 750 keV (which is the mean $E_{peak}$ in source frame if we use \=z = 0.53), is small (it depends on the detector and on the burst's redshift but is always $\lesssim$ 5\% , which can be understood by the fact that the spectrum is shallow (Band function alpha = -0.5)).} 
}
), $\alpha_{BAND}=-0.5$, $\beta_{BAND}=-2.25$. Note that this definition of the luminosity give luminosities which are larger by a factor of $3-10$ - depending on the redshift and on the detectors' energy window  - compared with the luminosity in the ``classical" energy band  \citep[$50keV-300keV$ as in e.g.][]{GP06}. The peak luminosity is related to $P$,  the 64-ms peak photon count  as:
\begin{equation}
 L = 4 \pi D(z)^2 (1 + z) k(z) C_{det} ~ P\ ,
 \label{L}
\end{equation}
where  $D(z)$ is the proper distance at redshift $z$\footnote{We adopt a standard $\Lambda CDM$ cosmology with $h_0=0.7$, $\Omega_{\Lambda}=0.7$ and $\Omega_m=0.3$.}. $C_{det}$ (having units of energy) is the total $\gamma$-ray luminosity divided by the number rate of photons in the detector's energy window from a source at redshift = 0:
\begin{equation}
C_{det} = \frac{\int_{1keV}^{10MeV}E N(E) dE }{\int_{E_{min}}^{E_{max}} N(E) dE} \ .
\end{equation}
Finally, $k(z)$ is the k-correction for the given spectrum at redshift z,
\begin{equation}
k(z) = \frac{\int_{E_{min}}^{E_{max}} N(E) dE}{\int_{(1+z)E_{min}}^{(1+z)E_{max}} N(E) dE} \ , 
\end{equation}
where $N(E)$ is the Band function and $[E_{min}, E_{max}]$ is the detector's energy window.

With the ns$^2$ merger model in mind we construct a rate function that describes a rate that has a time delay (corresponding to the spiral-in time of the binaries) relative to the global SFR. 
Such a rate is given as  a convolution of a given selected  SFR (we consider in the following two SFR functions) and a time delay, $\Delta t$, with a  distribution, $f(\Delta t)$,  relative to the SFR. We stress that while motivated by the ns$^2$ model we allow for the possibility of no time delay at all and let the maximal likelihood procedure  find the best fit values. Thus, we do not force the ns$^2$ model on the data. 
The intrinsic sGRB rate is a convolution of the SFR with $f(\Delta t)$:
\begin{equation}
 R_{sGRB}(z) \propto \int_z^{\infty}{SFR(z')(f(t(z)-t(z'))\frac{dt}{dz'}dz'} \  .
\label{eqn:rates}
\end{equation}

We consider two models for the time delay:\\
I. A power-law time delay with index $\alpha_t$ and a minimum time delay of $20Myr$: $f(\tau) = \tau^{-\alpha_t}$ for $\tau > 20 Myr$. Note that the simple ns$^2$ merger model  favors $\alpha_t=1$ \citep{Piran1992b}.  At the limit $\alpha_t \rightarrow \infty$, $R_{sGRB}$ tends to the SFR and the model is almost indistinguishable from the SFR  for $\alpha_t > 3$ .\\
II. A log-normal distribution with a width $\sigma_t$ around a  time delay $t_d$: $f(\tau) = \exp[{-\frac{(\ln\tau-\ln t_d)^2}{2\sigma_t^2}}]/({\sqrt{2\pi}\sigma_t})$. \\
For small values of $\sigma_t$ this function converges, of course, to a constant time delay. Hereafter we refer to that limit as the constant time delay model. In some cases we will consider this case in the analysis. 

\begin{figure}
 \includegraphics[width=\linewidth]{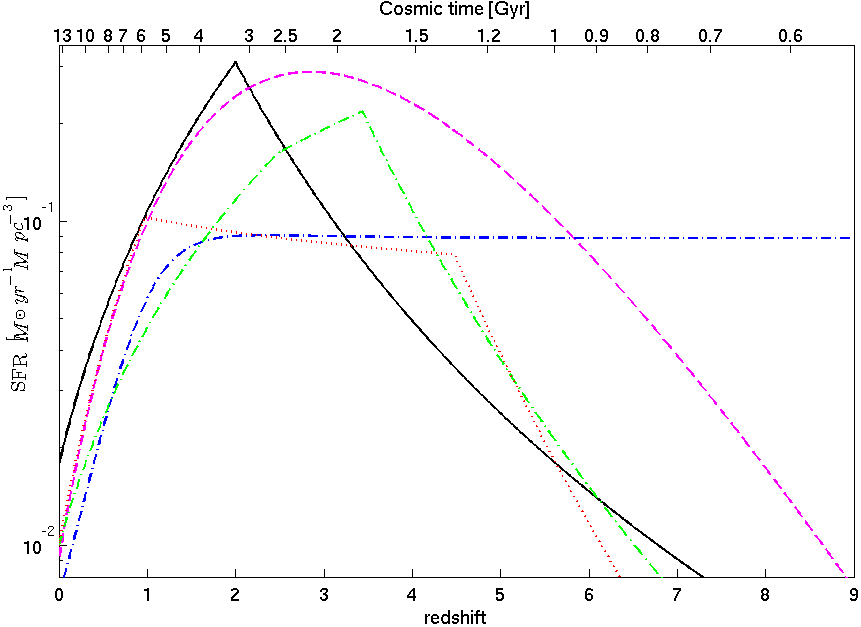}
 \caption{The star formation rate models vs. redshift. \protect\cite{Cucciati2012a} (SFR1) - black solid line, 
\protect\cite{PlanckCollaboration2013} Halo model (SFR2) - magenta dashed line. Three other commonly used SFRs which we do not use are plotted here  to emphasize the range of  SFRs in the literature: \protect\cite{Porciani2001} SF2 model - blue dash-dotted line, \protect\cite{Bouwens2011b} - green dashed line (peaks at redshift $\sim 3.4$) and  \protect\cite{Hopkins2006} piecewise linear model - red dotted line (almost flat in the redshift range $1-4.4$).
}
 \label{fig:sfrs}
\end{figure}

We examine two SFR models: (i)  We combine the low redshift SFR of  \cite{Cucciati2012a} with the high redshift part ($z>4$) of \cite{Bouwens2012} and \cite{Oesch2013} (We denote this as SFR1). The lower redshift  part of this SFR is measured using the VIMOS-VLT Deep Survey (VVDS), a single deep galaxy redshift survey \citep{Cucciati2012a} up to redshift $z \sim 4.5$. The higher redshift part is derived from the UV LFs from the Hubble ultra-deep WFC3/IR data \citep[We used the compilation in][]{Oesch2013}.  
(ii) The  \cite{PlanckCollaboration2013} Halo model (denoted SFR2). This is based on a completely different method which estimates the SFR from the cosmic infrared background (CIB) anisotropies measured with Planck. This method is  based on a model that associates star-forming galaxies with dark matter halos and their sub-halos, using a parametrized relation between the dust-processed infrared luminosity and (sub-)halo mass. SFR2 is not expected to be very accurate for $z>2$, nevertheless these two SFR models represent a range of feasible SFR models which allow us to estimate the sGRBs rate time delay with respect to different SFRs.
The two SFR models are plotted in figure \ref{fig:sfrs}, together with few other commonly used  SFR models.   

\subsection{A two component analysis }
\label{sec:met_twocomp}
We have asserted, following \cite{Bromberg2013}, that \swift sGRBs are composed of two populations, one of genuine non-Collapsars and the other of ``impostor" namely short Collapsars. To explore and demonstrate this assertion  
we have incorporate another model that allows for two populations  that have different event rates.  The two populations have the same luminosity function, as the sample is not large enough to explore two different luminosity functions at the same time. One of the two populations has  a time delay relative to the SFR, while the other one follows the SFR with no delay. The  free parameter $f_{SFR}$ determines the relative fraction of these two components:
\begin{equation}
 R(z) = (1-f_{SFR})\cdot R_{sGRB}(z) + f_{SFR} \cdot SFR(z) ,
\end{equation}
where $R_{sGRB}(z)$ is given by equation \ref{eqn:rates}.
If our assertion is correct we expect that  when applying this rate model to the  full \swift sample  $f_{SFR}$ will obtain the fraction of non-Collapsar that is determined using other methods (combination of duration of hardness). This will demonstrate that indeed it is composed of two distinct populations { (see \S \ref{sec:results})}.

\subsection{The likelihood function}
\label{subsec:likelihood_function}
The likelihood function, $\cal L$, describes the combined likelihood of observing the BATSE, Fermi and \swift data:
\begin{equation}
\begin{split}
{\cal L}=\prod_{i}^{BATSE}\left(\frac{N'(P_{i})}{\int_{P_{min}^{BATSE}}^{P_{max}^{BATSE}}N'(P)dP}\right)
  \prod_{j}^{Fermi}\left(\frac{N'(P_{j})}{\int_{P_{min}^{Fermi}}^{P_{max}^{Fermi}}N'(P)dP}\right) \\
  \prod_{k}^{Swift}\left(\frac{\phi_0(L_{k})R_{sGRBs}(z_k)}{\int_{P_{min}^{Swift}}^{P_{max}^{Swift}}N'(P)dP}\right),
\end{split}
\end{equation}
where the product is over all bursts in the  BATSE (i), Fermi (j) and {\it Swift} (k) samples. 
The observable number density for a peak flux $P$  
is given by:
\begin{equation}
 N'(P)=\frac{1}{P}\int_0^\infty R_{sGRBs}(z)\phi_0(L(z,P))dz,
\label{eqn:Ntag_of_P}
\end{equation}
where $L(z,P)$ is given by equation \ref{L}. 

The best-fit parameters are found by maximizing the likelihood over the parameter space for each parameter or by marginalizing the likelihood over of the sub-space of all the other parameters. In all cases the best-fit parameters do not significantly differ whether maximizing or marginalizing the likelihood. The uncertainty range is estimated using the likelihood ratio where 
$1\sigma$ (68\%) uncertainty level corresponds to a likelihood ratio of $e^{-0.5}$. The above uncertainty level estimation is validated by a Bootstrap Monte-Carlo simulation. We note that this might be a poor estimator in cases where the maximum value of the likelihood is found at the edge of the parameter space as in the case of the width of the log-normal distribution for time delay.
The local event rate $\rho_0$ is found by comparing the total  expected number of events from the maximum likelihood model to the total observed BATSE and Fermi sGRBs and correcting for the non-Collapsars fraction of all burst with $T_{90}>2s$ following \cite{Bromberg2013}. 
By performing this calculation for all points in the parameter space for which the likelihood ratio is above $e^{-0.5}$ we 
get a range of  $\rho_0$ values which is the estimated $1\sigma$ uncertainty range of  $\rho_0$.

\section{Results}
\label{sec:results}

\begin{figure}
\begin{subfigure}{.48\textwidth}
  \includegraphics[width=1\linewidth]{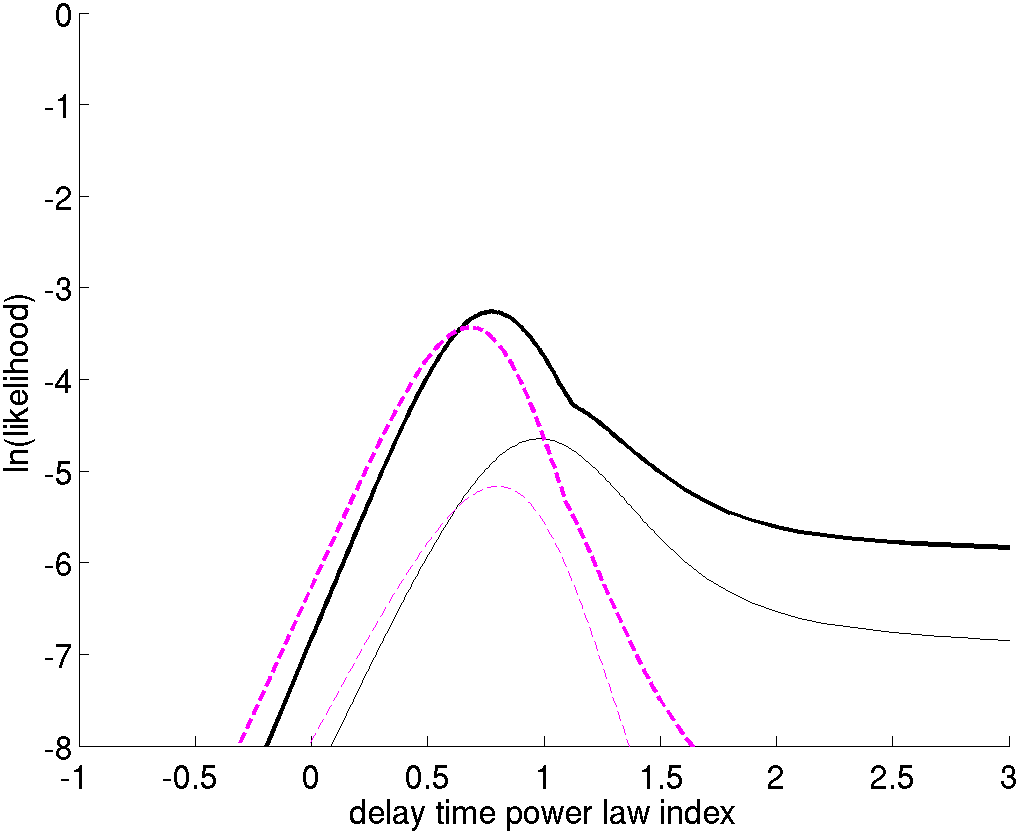}
\end{subfigure}
\begin{subfigure}{.48\textwidth}
  \includegraphics[width=1\linewidth]{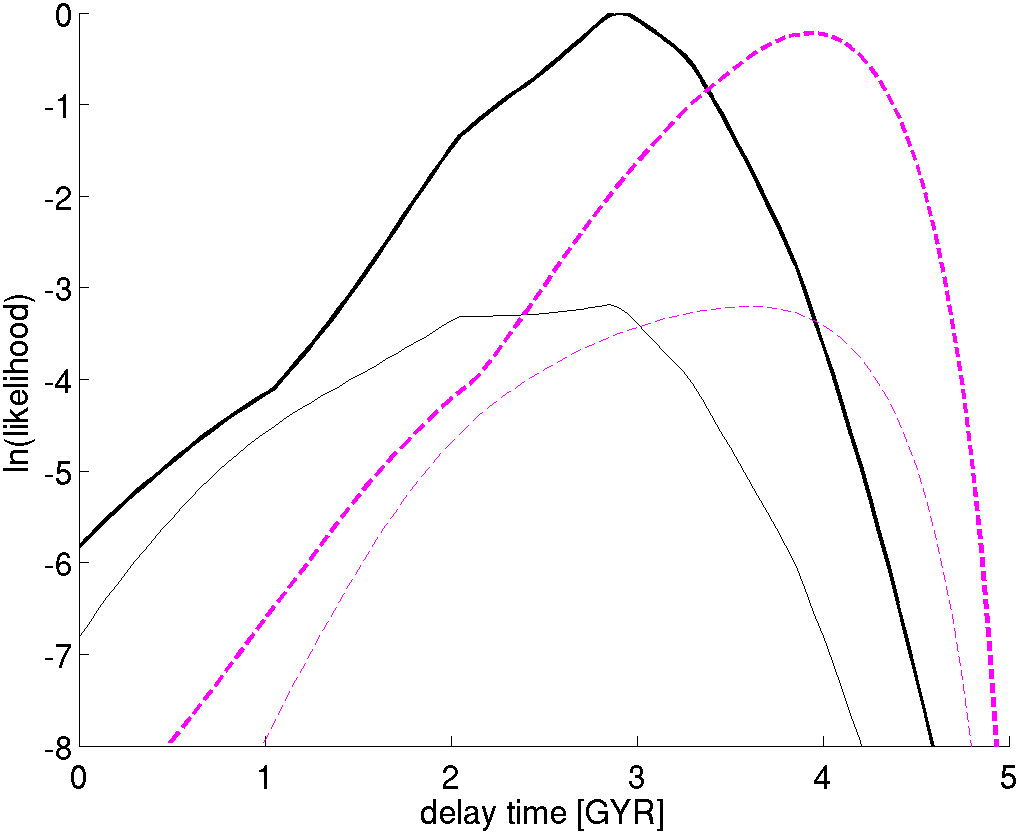}
\end{subfigure} 
\caption{The maximum likelihood as a function of the power law time delay parameter $\alpha_t$ (Top panel) / constant time delay parameter $t_d$ (Bottom panel) - all are normalized by the maximum likelihood of the highest model. The line color and shape denote the SFR model: SFR1 - black solid line, SFR2 - magenta dashed line. Each of the SFRs have two lines in each panel: the upper-thick line is for a broken power-law luminosity function model and the lower-thin line is for the single power-law luminosity function model.  }
 \label{fig:maxs_pltd_and_ctd}
\end{figure}

\begin{figure*}
 \includegraphics[width=1\linewidth]{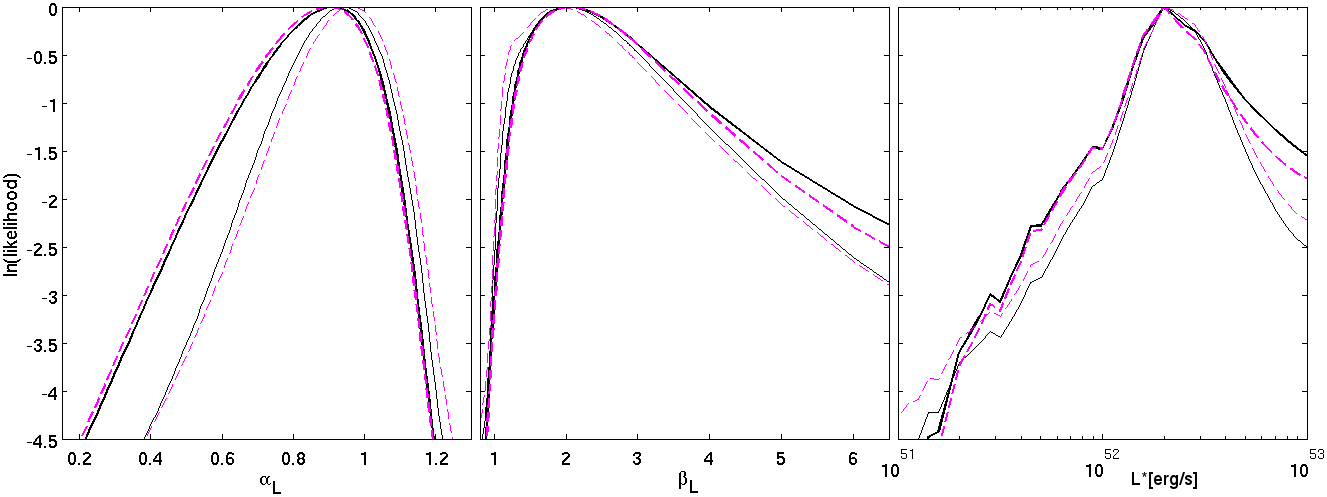}
 \caption{The marginalized likelihood ratio as a function of each the luminosity function parameters $\alpha_L$, $\beta_L$ and $L*$ (for the three panels from left to right). 
 The line color and shape denote the SFR model: SFR1 - black solid line, SFR2 - magenta dashed line. Each of the SFRs have two lines: the thin line is for a power-law time delay model and the thick line is for the constant time delay model (each model is normalized separately). }
 \label{fig:mgz_abL}
 \end{figure*}

\begin{figure*}
 \includegraphics[width=1\linewidth]{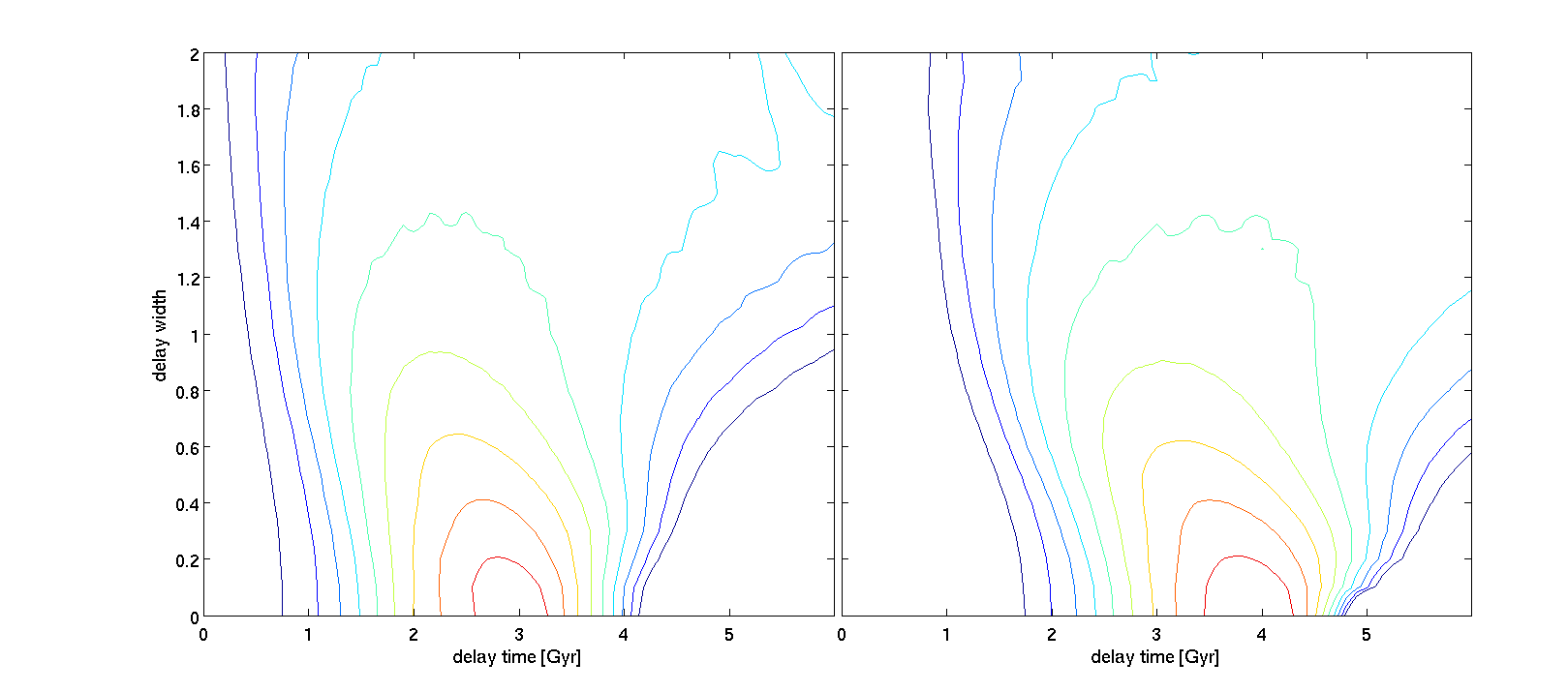}
\caption{Contours of the log-likelihood as a function of the time delay in Gyr and the standard deviation of the time delay log-normal distribution.
Left hand side panel: SFR1, right hand side panel: SFR2. The contour levels represent a factor of $e^{0.5}$ in likelihood between two adjacent lines.}
 \label{fig:ctdw}
 \end{figure*}

\begin{figure}
 \includegraphics[width=1\linewidth]{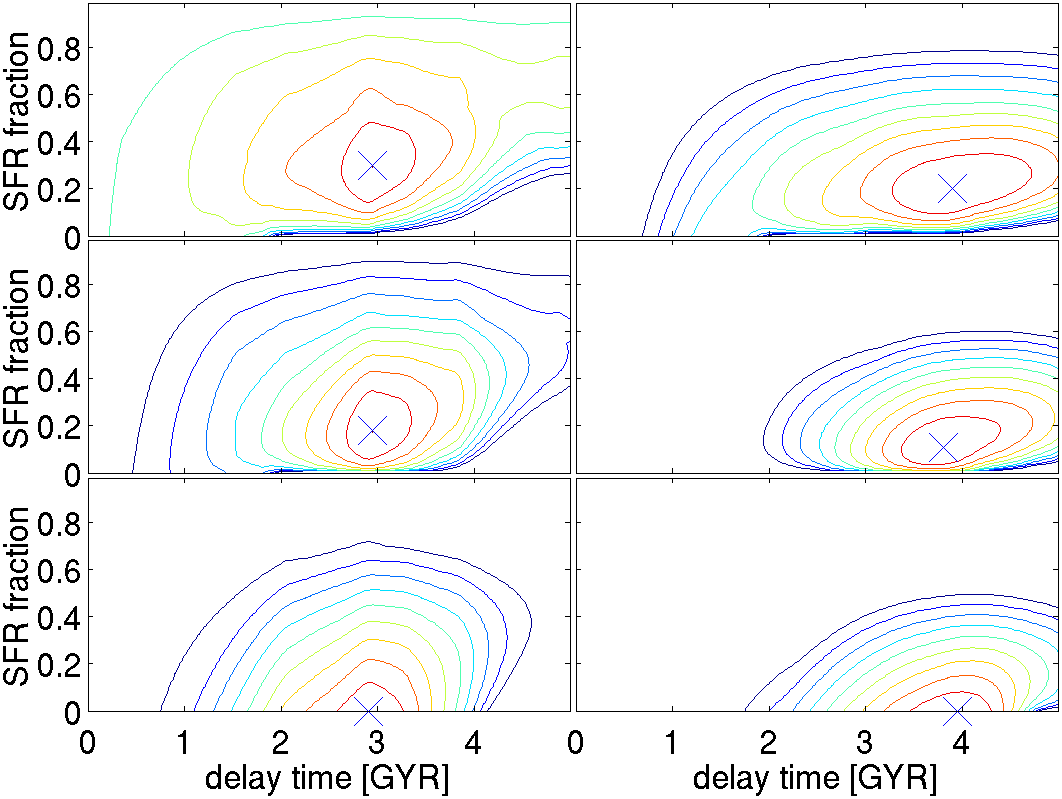} 
 \caption{Contours of the maximum log-likelihood as a function of the time delay parameter and the SFR fraction for the 'two time delay components' model. The different panels are for the two SFR models we study and for the sub-samples of bursts with redshifts: Left hand side panels: SFR1, right hand side panels: SFR2, top and middle panels: for all 20 bursts for burst 050813 at z=1.8 and z=0.722 respectively, bottom panels: for the 12 bursts with $f_{NC}>0.6$.
 The contour levels represent a factor of $e^{0.5}$ in likelihood between two adjacent lines.}
 \label{fig:td_a}
\end{figure}

Our results are shown in Figures \ref{fig:maxs_pltd_and_ctd} - \ref{fig:td_a}.
Figure \ref{fig:maxs_pltd_and_ctd} depicts the maximum likelihood as a function of 
the delay power-law index or of the time delay (the latter for the constant time delay model). Figure \ref{fig:mgz_abL} depicts the likelihood for each of the parameters of the broken power-law luminosity function.
Figure \ref{fig:ctdw} depicts the likelihood contours in the log-normal distribution parameter space (delay time $t_d$ - delay width $\sigma_t$) and Figure \ref{fig:td_a} depicts the likelihood contour lines for the SFR fraction parameter, $f_{SFR} $, of the two component model (both for the full sample analysis and for the non-collapsars sub-sample).
We present the results as curves of the likelihood as a function of different parameters. 
Figure \ref{fig:maxs_pltd_and_ctd} depict the maximal likelihood for the given parameter shown in the figure where we scan the phase-space of the rest of the model parameters. This enables us to compare different parameter spaces, i.e. a single power-law and a broken power-law models for the luminosity function.
Figures \ref{fig:mgz_abL} - \ref{fig:td_a} depict the marginalized likelihood for the given parameter(s) shown in the figure over the phase-space of the rest of the model parameters. 
The results of the maximal likelihood in all cases are very similar to the results of marginalizing the likelihood, thereby confirming the robustness of the result.
The most likely value and uncertainty range (after marginalization) for each parameter are summarized in Table \ref{tbl:results}.  Also reported in Table \ref{tbl:results} are the likelihood ratios, namely  the ratios of the maximal likelihood in a given  model and the maximal likelihood among all four models examined. This enables us to compare between the different models.

The likelihoods for the single power-law luminosity function and the broken power-law luminosity function are shown in figure  \ref{fig:maxs_pltd_and_ctd}. The likelihood for the broken power-law model is higher than likelihood for the single power-law model by a factor $3-4$ which corresponds to a rejection of the single power law model at  $\sim 2.5\sigma$. We adopt, therefore, the more general broken power-law luminosity function for the rest of the analysis.

Figure \ref{fig:maxs_pltd_and_ctd} depicts the likelihood for the two SFRs and for the two time delay distributions.
For both SFR models when considering the power-law time delay distribution  $\alpha_t$, the time delay power law index peaks around $0.7-0.8$. This is  consistent with  the  value 1, expected for a distribution that arises 
from mergers \cite{Piran92}. 
However for both SFR models the maximal likelihood of the constant time delay model (or the log-normal distribution) was significantly higher than the maximum likelihood of the power-law time delay model.
This means that the power law delay time model is disfavoured as describing the rate of sGRB.
The power-law delay time model predicts a significant fraction sGRBs with short delays and hence a significant fraction of $z\gtrsim 1$ bursts while the log-normal time delay model successfully suppress the $z\gtrsim 1$ population and is consistent with  the observed redshift distribution. From all our results this result, which has important physical implications that we discuss later, is most sensitive to observational bias. In particular this conclusion strongly depends on the observed redshift sample and it depends critically on the lack of $z>1.2$ bursts in our non-Collapsar sGRB sub-sample.  
Note that among the likely Collapsars sGRBs that we have eliminated from our sample there are a few bursts above this redshift. 
A detection of two or three non-Collapsar sGRBs with a higher redshift may change this conclusion.

The likelihood of the constant time delay model peaks at $2.9$~Gyr and $3.9$~Gyr for SFR1 and SFR2 respectively. 
This time delay ranges from $2.5$~Gyr to $3.3$~Gyr for SFR1 and from $3.4$~Gyr to $4.3$~Gyr for SFR2.
Figure \ref{fig:ctdw} depicts the likelihood contours for the log-normal model as a function of the average time delay and the spread around this average. The likelihood peaks (for both SFRs) at zero width i.e. at a constant time delay. The width of the log-normal distribution is $\sigma_t \lesssim 0.2$ (at 68\% confidence), which corresponds to a very small spread by a factor of $\lesssim 1.2$ in the time delays. As a constant time delay gives the best fit we adopt it throughout the paper, replacing the more general log-normal distribution.

{  As mentioned earlier it is possible that some burst at $z>1.2$ are missed because redshift measurements in this range are harder, as their main spectral lines leave the optical band. To study the effect of such bias on the main result of the paper - the preferred time delay model - we have simulated sets of redshift samples by adding a varying number of bursts uniformly distributed in the redshift range $1<z<2$ to our observed sample of 12 non-collapsars sGRBs, and have randomly drawn the luminosity from the luminosity function we get for the real sample, taking only the part which is above the detector's threshold for the given redshift. We have repeated the analysis of the resulted likelihood function for each simulated redshift sample and specifically looking on the likelihood ratio between the log-normal and the power-law time delay models.
The conclusion is that by adding 2-3 bursts in the $1<z<2$ redshift range we void the preference of the log-normal time delay model making both model consistent with the simulated samples. Adding further more bursts at that redshift range does not change this, as the parameters of each of the time delay models vary but they are both consistent as no model can be said to be significantly more likely than the other.
}

Once we have figured out the parameters for the ``uncontaminated" \swift sample we turn  to examine the whole sample of \swift sGRBs with redshifts  (still excluding sGRBs with long soft tails). In this case, as mentioned earlier in \S \ref{sec:met_twocomp}, we have modeled the sGRB rate as as sum of two components, one with a constant time delay and the other with no time delay relative to the SFR. The ratio of these two components is not set a priori and it is a new free parameter, which we denote as the ``SFR fraction" $f_{SFR}$.  We carry out this analysis twice. Once,
on the full \swift sample (20 bursts) and once as a check on the uncontaminated sample  from which we have kept only \swift sGRBs that have high probability  of being Collapsars (12 bursts). The full sample itself has two variants because the redshift of burst 050813 may be either 1.8 or 0.722. These two options are  referred to as ``full sample (A)" and ``full sample (B)" respectively.

Figure \ref{fig:td_a} depicts contours of the  likelihood as a function of the time delay parameter $t_d$ and the ``SFR fraction" $f_{SFR}$ for the two SFR models that we consider and for the two samples. The best-fit values and the $1\sigma$ uncertainty ranges are summarized in Table \ref{tbl:results_fsfr}.  As expected the maximal likelihood for the uncontaminated sample arises when all bursts have a delay (the ``SFR fraction" vanishes) and in this case we recover the delay times derived earlier. On the other hand once we consider the full sample we find that the maximal likelihood is obtained for $f_{SFR} = 0.30_{-0.15}^{+0.18}$ for the full sample (A) and $f_{SFR} = 0.18_{-0.12}^{+0.17}$ for the full sample (B) with SFR1 and $f_{SFR} = 0.20_{-0.10}^{+0.15}$ for the full sample (A) and $f_{SFR} = 0.11_{-0.07}^{+0.13}$ for the full sample (B) with SFR2. 
This number is to be compared with the fraction of Collapsars in the full sample which can be read from the last line in Table \ref{tbl:sGRBs_sample} $f_C = f_{Collapsar} = 1 - f_{Non-Collapsar}$ =  0.34. The $f_{SFR}$ values for both SFR models are significantly inconsistent with zero but they are consistent with $f_C$ of the full sample,  providing an independent evidence that \swift short GRBs contain a significant fraction of short Collapsars  \citep{Bromberg2013}. 
One cannot rule out, of course, the possibility that all bursts are genuine non-Collapsars and that there is a population of non-Collapsars sGRBs that follows the SFR without time delay. Such a possibility has been found in some  population synthesis models \citep{belczynski02b}. 

\begin{figure}
\begin{subfigure}{.48\textwidth}
  \includegraphics[width=1\linewidth]{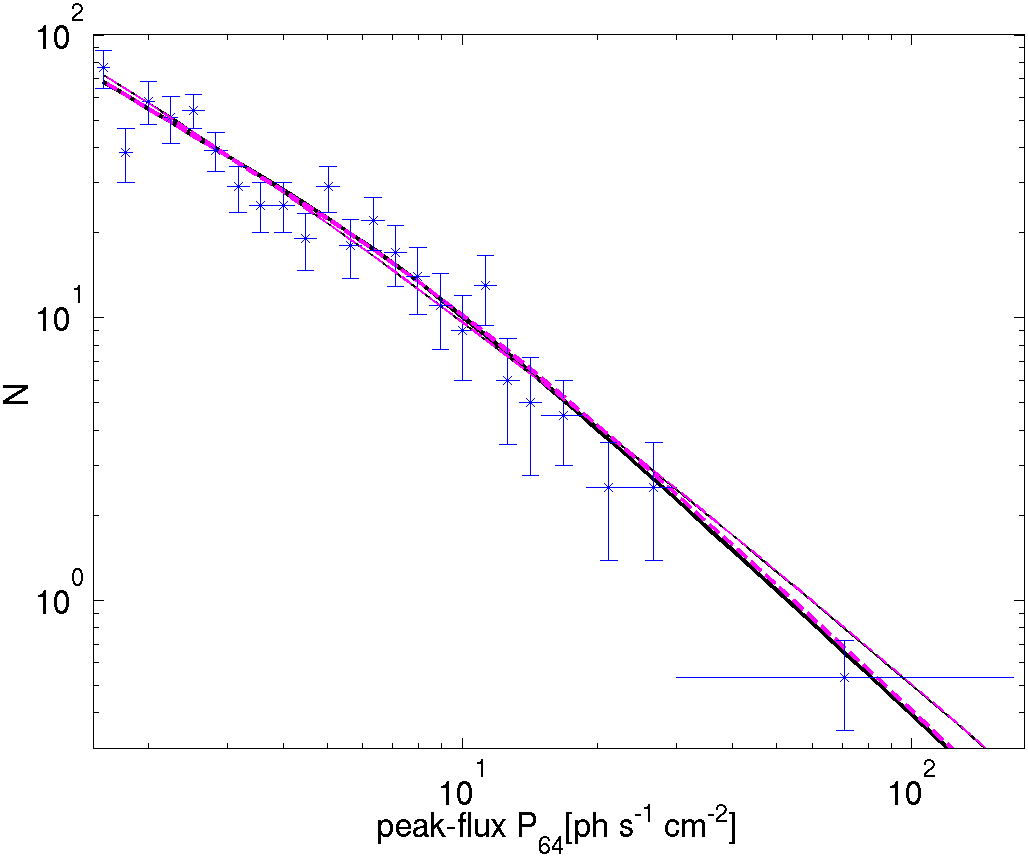}
\end{subfigure}
\begin{subfigure}{.48\textwidth}
  \includegraphics[width=1\linewidth]{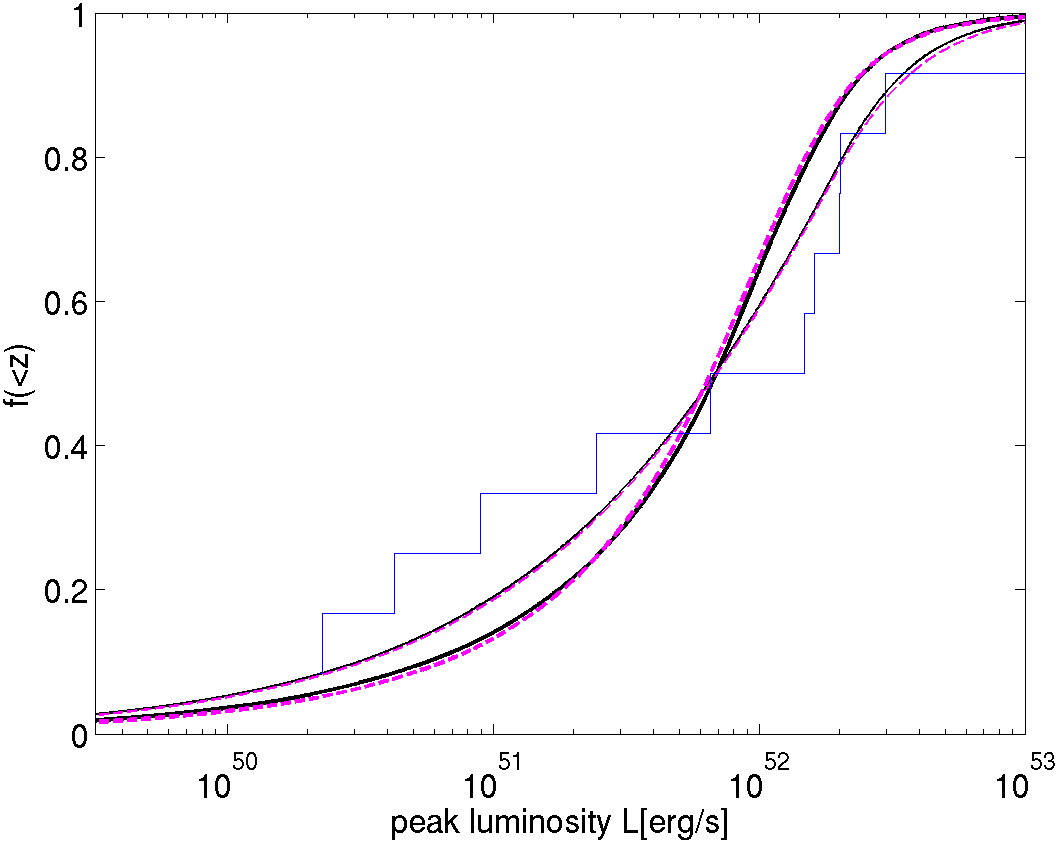}
\end{subfigure} 
 \caption{{\it Top panel:} peak-flux distribution and the models found for BATSE and Fermi together (see text for details). The modeled peak-flux distributions are consistent with the observed sample: $\chi^2=19.4, 19.7, 21.0$ and $21.0$ (d.o.f.=23) for the log-normal time delay model with SFR1 or SFR2, and the power-law time delay model with SFR1 or SFR2, respectively.
 {\it Bottom panel:} the observed (12 bursts with $f_{NC}>0.6$) and the predicted luminosity cumulative distribution for the log-normal time delay model (thick lines) and the power-law time delay model (thin lines). 
 For both the peak-fluxes and the luminosity distributions, the black solid line and the magenta dashed line corresponds to SFR1 and SFR2 respectively.
 The modeled luminosity distributions are not inconsistent with the observed sample: $P_{KS}=0.24, 0.19$ for the log-normal time delay model with SFR1, SFR2 and $P_{KS}=0.64, 0.67$ for the power-law time delay model with SFR1, SFR2.}
 \label{fig:best_fit_fits}
\end{figure}

\begin{figure}
\begin{subfigure}{.48\textwidth}
  \includegraphics[width=1\linewidth]{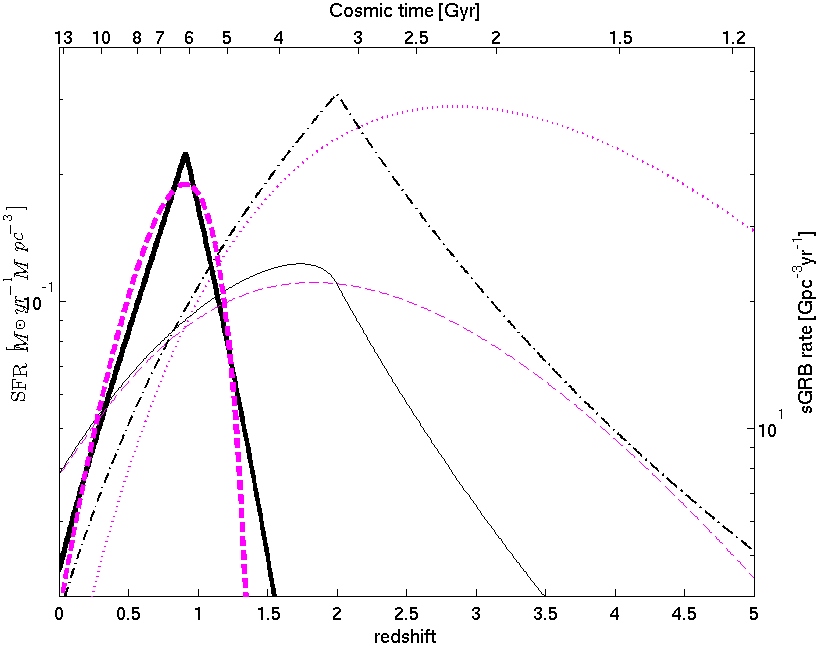}
\end{subfigure}
\begin{subfigure}{.48\textwidth}
  \includegraphics[width=1\linewidth]{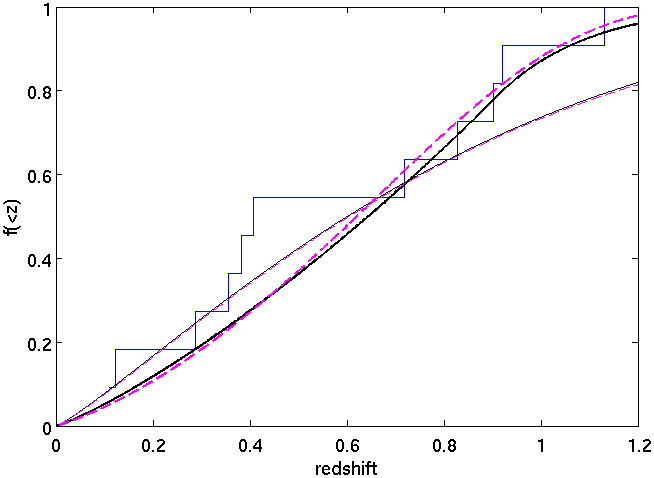}
\end{subfigure} 
 \caption{{\it Top panel:} The SFRs - (black dashed-dotted and magenta dotted lines for SFR1 and SFR2 respectively) and the sGRB rates for the log-normal time delay models (thick solid and thick dashed lines) and the power-law time delay model (thin solid and thin dashed lines). 
 {\it Bottom panel:} The observed (12 bursts with $f_{NC}>0.6$) and the predicted redshift cumulative distribution for the log-normal time delay model (thick lines) and the power-law time delay model (thin lines). 
 For both panels, the black solid line and the magenta dashed line corresponds to SFR1 and SFR2 respectively.
 The modeled redshift distributions are not inconsistent with the observed sample: $P_{KS}=0.36, 0.34$ for the log-normal time delay model with SFR1, SFR2 and $P_{KS}=0.63, 0.62$ for the power-law time delay model with SFR1, SFR2.
 }
 \label{fig:SFRs_and_ctd_rates}
\end{figure}

As a demonstration of the quality of the fit  we compare the predictions of the model for BATSE, Fermi and \swift with the observations. 
Figure \ref{fig:best_fit_fits} and Figure \ref{fig:SFRs_and_ctd_rates} compare the observed samples of BATSE, Fermi and \swift with the best-fit models 
of both a constant time delay and a power law time delay with SFR1 and SFR2\footnote{We divide the data to bins equally spaced in logP, then, to improve statistical accuracy, we merge adjacent bins which have less than 5 bursts.}. For the peak-flux distribution we have used the effective full sky observing time of BATSE and Fermi i.e. observations period times the field of view (4.44 yr for BATSE and 3.65 yr for Fermi), and corrected accordingly the first four bins with peak-flux $<2.37 ph~cm^{-2}s^{-1}$ where we have only BATSE bursts in our sample.
We obtain good fits for the models (we denote here the results corresponding the two SFR models in the format [SFR1;SFR2]) with $\chi^2/d.o.f. = [21.4/19;21.4/19], [15.3/12;15.5/12]$ and $[19.4/23;19.7/23]$ for BATSE, Fermi and both, respectively for the constant time delay model and $\chi^2/d.o.f. = [22.7/19;22.7/19], [16.7/12;16.6/12]$ and $[21.0/23;21.0/23]$ for BATSE, Fermi and both, respectively for the power law time delay model. 
We have also tested the luminosity and redshift distributions of our Swift sample using a Kolmogorov-Smirnov (KS) test. For the constant time delay model
we find  a  KS  probabilities\footnote{  A model is considered rejected for probabilities lower than some threshold, usually $P_{KS}  <$ 0.05. All higher values cannot be rejected and preferring one over the other is not statistically significant.} of $[0.24;0.19]$ and $[0.36;0.34]$ for the luminosity and redshift distributions respectively. We find KS values of  $[0.64;0.67]$ and $[0.63;0.62]$ for the luminosity and redshift distributions respectively for the power law time delay model.

\begin{table}
\begin{tabular}{|l|llll|}
\hline
time delay model & log-normal & log-normal & power-law	& power-law	\\
SFR model & SFR1	& SFR2	& SFR1	& SFR2	\\
\hline
likelihood ratio &  1	& 0.801	& 0.038	& 0.032	\\
\hline
$\rho_0$ [$Gpc^{-3}yr^{-1}$] & $4.6_{-1.7}^{+1.9}$	& $3.6_{-1.4}^{+1.6}$	& $7.8_{-4.5}^{+5.1}$	&  $7.7_{-4.6}^{+5.4}$	\\
$\alpha_L$ & $0.94_{-0.13}^{+0.11}$ & $0.96_{-0.12}^{+0.11}$ & $0.91_{-0.17}^{+0.11}$ & $0.90_{-0.17}^{+0.12}$ \\ 
$\beta_L$ & $2.0_{-0.7}^{+1.0}$ & $1.9_{-0.7}^{+1.0}$ & $2.0_{-0.6}^{+1.1}$ & $2.1_{-0.7}^{+1.0}$ \\ 
$L* $ [$10^{52}$ erg/s]& $2.0_{-0.4}^{+1.3}$	& $2.0_{-0.4}^{+1.4}$	& $2.0_{-0.5}^{+1.5}$	& $2.0_{-0.5}^{+1.2}$	\\ 
\hline
$t_d$ $[Gyr]$& $2.9_{-0.4}^{+0.4}$	& $3.9_{-0.5}^{+0.4}$	& & \\ 
$\sigma_t$ & $0^{+0.2}$	& $0^{+0.2}$	& & \\ 
\hline
$\alpha_t$ &	&	& $0.81_{-0.24}^{+0.25}$ & $0.71_{-0.23}^{+0.21}$ \\
\hline
\end{tabular}
\caption{The parameters best-fit and $1\sigma$ uncertainty levels for the log-normal time delay and the power-law time delay and for the different SFR models.}
\label{tbl:results}
\end{table}

\begin{table}
\begin{tabular}{|cc|lll|}
\hline
Model & 	 &  Full sample (A) & Full sample (B) & non-Collapsars sample\\
\hline
SFR1 	& log-normal	& $0.34_{-0.19}^{+0.27}$	& $0.18_{-0.12}^{+0.22}$	& $0^{+0.12}$ \\
SFR2 &	log-normal & $0.23_{-0.13}^{+0.22}$	& $0.11_{-0.07}^{+0.16}$ 	& $0^{+0.08}$ \\
\hline
\end{tabular}
\caption{$f_{SFR}$ (the fraction of bursts which follows the SFR) for the full \swift sample and for the reduced, non-Collapsar sample. Full sample A and B refer to the sample of swift sGRBs with burst 131004A at redshifts 1.8 and 0.722 respectively.}
\label{tbl:results_fsfr}
\end{table}

\section{Discussion}
\label{sec:discussion}
{  We have carried out a joint analysis of the BATSE, Fermi and \swift sGRB data to determine the luminosity function and the rate of these bursts. It turns out that the rate is determined mostly by the \swift data (that has a redshift) while the BATSE-Fermi data is essential to determine the luminosity function. The combined data set give, of course better constraints on both. } 
We have found best fit parameters for the luminosity function and the rate of sGRBs. We focused on a sub-group of the \swift sGRB sample that has a high probability for being non-Collapsars. We have also considered the full sample and found evidence that it is composed of two distinct populations, in agreements with the expectation that this sample includes both non-Collapsar and Collapsars with  short durations. 

The (logarithmic) luminosity function is best fitted with a broken power-law with a break at luminosity of $L*=2 \times 10^{52} $~erg/s and with indices of $\alpha_L \simeq 1$ and $\beta_L \simeq 2$ for all the SFR and time delay models we have studied.
These  power-law indices are consistent with those found in  previous studies \citep[e.g.][]{GP06,GS09}.
However, the break luminosity is higher compared with these previous studies. This is largely due to the fact that we use here  the luminosity within $ 1-10000 $~keV whereas previous studies typically consider the luminosity within  $ 50-300 $~keV (see \S\ref{subsec:the_models}).
\cite{Nakar2006}  have found a single power-law luminosity function with power-law index $\simeq 1 $. We find that while the low-end power-law is consistent with this power-law, our high-end power-law  deviate significantly from it,  introducing an upper drop in the luminosity function. 
When comparing the luminosity function found here for sGRBs  with the long GRBs luminosity function  \citep{WP10} we find that the break luminosity is close ($L*_{sGRB}=2 \times 10^{52}$ and $L*_{LGRB}=3 \times 10^{52}$), but the sGRB luminosity function is steeper compared with the LGRB luminosity function (both the low-end and the high-end power-law indices of the sGRB luminosity function are larger by 0.6 compared with the LGRB luminosity function power-law indices).

The log-normal time delay model (actually its limit of a constant time delay) is favored compared with the power-law time delay model for all the SFRs we have examined. In cases when we considered a log-normal time delay distribution we found that the width of the distribution  seems to be rather narrow as the likelihood peaks at a zero width and at a significance level of $95\%$,  $\sigma_t< 1$ (corresponding to variation of about a factor of 2 in the time delay). We stress again that among all the results this specific one depends most critically on the lack of $z>1.2$ redshift non-Collapsar sGRBs, 
{  and addition two or three high redshift bursts is sufficient to make the difference in likelihood of these time delay models statistically insignificant}.

The likelihood is maximal at a constant delay time parameter $t_d = 2.9^{+0.4}_{-0.4}$~Gyr and $t_d = 3.9_{-0.5}^{+0.4}$~Gyr for SFR1 and SFR2 respectively.
Since a long constant delay ($t_d > 5$~Gyr) is rejected by at least $4\sigma$ for all of our models, the time delay cannot be too long. This arises from the fact that the SFR  drops rapidly with redshift at redshift beyond $z \simeq 5$ while the highest redshift  sGRB that was observed is at $z = 1.131$ about $4 $~Gyr after redshift $z=5$.  The same observation shows that  
the time delay cannot be too short  
since  the bulk of star formation activity peaks at redshift $z\gae  2$ at least $2.5 $~Gyr before $z=1.131$. Indeed, the zero-time-delay model in which the sGRB rate directly follows SFR is rejected at a $3.5\sigma$ and a $4.2\sigma$  level for SFR1 and SFR2 respectively. 
The constant time delay sGRB rate model for both SFR models converges to a common shape which peaks at redshift $z \simeq 0.9$ and has a width of $\Delta z \approx 0.4 ; 0.6$ for SFR1;SFR2 respectively. The peak rate (at $z=0.9$) is a factor of $\simeq 10$ higher than the current ($z=0$) rate (see Figure \ref{fig:SFRs_and_ctd_rates}). The effective sGRB rate for the model of constant time delay with respect to SFR1 can also be described directly as a function of the redshift, rather than as a delay with respect to  the  SFR, as:
\begin{equation}
 R_{sGRB}(z) = 45_{[Gpc^{-3}yr^{-1}]} \times
\left\{
\begin{array}{ll}
 e^{(z-0.9)/0.39}   & z \leq 0.9 \ , \\
 e^{-(z-0.9)/0.26} & z > 0.9 \ .
\end{array}
\right. 
\end{equation}

The estimate of the local rate of sGRBs depends critically on the lower end of the luminosity function.
Due to the limited number of sGRBs with redshift observed by \swift, we cannot determine the cutoffs of the power-law luminosity function at either the low or the high end. 
{  Given that BATSE is more sensitive to short GRBs relative to \swift, we can expect that the lowest luminosity burst of the BATSE sample, to which we calibrate the overall rate is smaller than the least luminous burst in our \swift sample.
}
We choose to adopt here the value $5 \times 10^{49}$~erg/s as our 'canonical' value, elaborating later in great detail on the effect of varying this lowest luminosity limit. The total current rate is therefore: 
\begin{equation} 
 \rho_0= (4.1_{-1.4}^{+1.8}\pm 0.5) ~f_b^{-1} (\frac{L_{min}}{ 5 \times 10^{49}~{\rm erg/s}})^{-0.95} {\rm  Gpc}^{-3} {\rm yr}^{-1} ,
\end{equation}
for the preferred time delay model, i.e the log-normal distribution, where the $\pm 0.5$ represent the difference in results between the models for SFR1 and SFR2. The beaming factor $f_b^{-1}$ is the fraction of bursts that point toward us due to the opening angle of the emitting jet beam. The beaming angle is in the range $1 < f_b^{-1} \lesssim 100$ \citep[][]{Nakar2007b, Fong2012}.

When comparing with previous studies 
\citep{Ando2004bi,GP05,GP06,Nakar2006,GS09,Dietz2011i,Coward2012b,Siellez2013}
we have to consider the fact that those studies used different lower cutoff to their sGRB luminosity functions. 
When doing so one has to recall that different authors have used different definitions of the luminosity and these have to be normalized to ours. Table \ref{tbl:results_comparision} summarized different earlier results
and compare them with ours. In most cases the results are within a factor of 2. 

\begin{figure}
 \includegraphics[width=\linewidth]{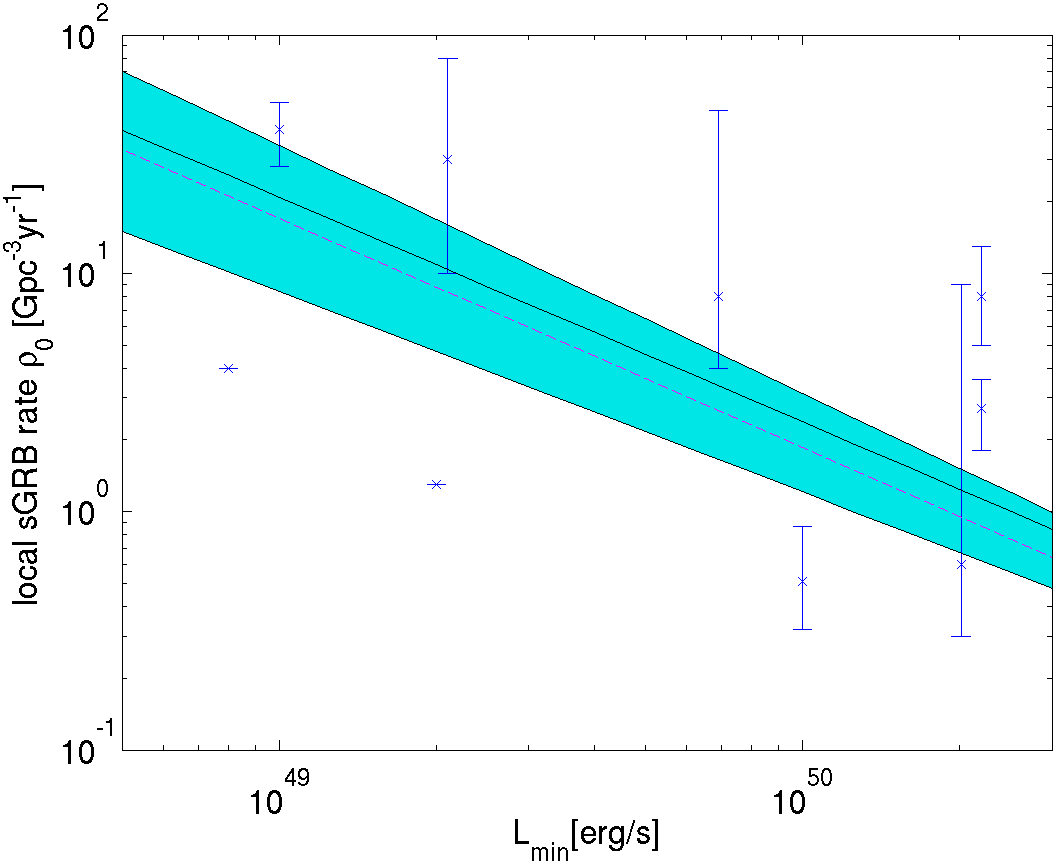}
 \caption{The local sGRB rate as a function of the minimal luminosity for SFR1 (black solid line) and SFR2 (magenta dashed line) for the best-fit parameters in the log-normal time delay model. The shaded area is the joint $1\sigma$ uncertainty range of the local sGRB rate for both SFR1 and SFR2 based models. The error bars represent the results of previous studies - see table \ref{tbl:results_comparision} for details.}
 \label{fig:rho_0_prev_studies}
\end{figure}

\begin{table}
\begin{tabular}{llll}
\hline
& local rate ($\rho_0$)  & $L_{min} $  & This work \\ 
& [Gpc$^{-3}$yr$^{-1}$] & [erg/s] & for a similar $L_{min}$ \\
\hline
\cite{Ando2004bi} (1)	& $0.51^{+0.36}_{-0.19}$ 	& $10^{50}$ & $2.1_{-0.9}^{+1.0}$ \\
\cite{GP06} (1)		& $0.6^{+8.4}_{-0.3}$ 	& $2 \times 10^{50}$ & $1.1_{-0.4}^{+0.4}$ \\
\cite{GP06} (2)		& $8^{+40}_{-4}$ 	& $7 \times 10^{49}$ & $3.0_{-1.4}^{+1.6}$ \\
\cite{GP06} (1)		& $30^{+50}_{-20}$ 	& $2 \times 10^{49}$ & $9.4_{-4.9}^{+6.6}$\\
\cite{Nakar2006} 	& $40^{+12}_{-12}$ 	& $10^{49}$ & $18.9_{-10.5}^{+15.5}$ \\
\cite{GS09} (1)		& $1.3$ 	& $2 \times 10^{49}$ & $9.8_{-5.1}^{+6.9}$ \\
\cite{GS09} (3)		& $4 $ 	& $0.8 \times 10^{49}$ & $23.4_{-13.2}^{+20.0}$ \\
\cite{Dietz2011i}	& $1.05^{+0.5}_{-0.9}$ 	& $4 \times 10^{50}(*)$ & $0.5_{-0.2}^{+0.2}$ \\ 
\cite{Coward2012b} 	& $8^{+5}_{-3}$ 	& $2 \times 10^{50}(*)$ & $1.0_{-0.4}^{+0.4}$ \\
\cite{Siellez2013} 	& $2.7^{+0.9}_{-0.9}$ 	& $2 \times 10^{50}(*)$ & $1.0_{-0.4}^{+0.4}$ \\
This work 	& $4.1_{-1.9}^{+2.3}$ 	& $5\times 10^{49}$ & $4.1_{-1.9}^{+2.3}$ \\
\hline
\end{tabular}
\caption{A comparison of different estimates of the local short bursts rate, without a beaming correction. \newline Notes: (*) When no luminosity low-end cutoff (erg/s) is specified we take the cutoff to be just  below the least luminous burst in the given sample. 
(1) For a model with time delay $1/t$ for $t>20Myr$ with respect to \protect\cite{Porciani2001} SF2. 
(2) For a model with a constant rate at all redshifts. (3) For a constant time delay of 6 Gyr with respect to \protect\cite{Porciani2001} SF2.}
\label{tbl:results_comparision}
\end{table}

A comparison with other works considering the time delay with respect to the SFR cannot be done directly since other papers use different SFR models. Many older papers use the \cite{Porciani2001} SF2 model which has been disfavored by later observations.
There is however an agreement that sGRBs do not follow the SFR directly i.e. without delay. 
\cite{Ando2004bi}, \cite{GP06}, \cite{GS09}, favor a rate model which is delayed with respect to \cite{Porciani2001} SF2.
\cite{Hao2013cv} adopt an hierarchical structure formation model from \cite{Pereira2010ax}, in which the SFR is derived using a Press-Schechter like formalism. That SFR peaks at $z \approx 3$ and then drop very slowly as the redshift increases (by a factor $\sim 4$ form $z=3$ to $z=10$), more resembling the \cite{Porciani2001} SF2 SFR rather than the SFRs we consider here.
Consistently with  our results, they found that the power-law time delay model is disfavored and that the log-normal time delay distribution with $\tau \gtrsim 3$ Gyr is consistent with the data.

In the frame-work of ns$^2$ or a nsbh mergers, the  narrow  time delay distribution  implies an extremely  narrow distribution of the initial separation between the two compact objects. This time delay is  less sensitive to the other parameters, like masses or eccentricities and the allowed spread in the time delay is consistent with the expected variation in these parameters. 
For canonical 1.4 solar masses neutron stars and circular orbits the initial separation are 
\citep{{Shapiro1983}}
$2.0\times 10^{11}$~cm and $2.2 \times 10^{11}$~cm for $2.9$~Gyr and $3.9$~Gyr respectively (the best time delay values for SFR1 and SFR2 respectively). 
These values vary very little at the allowed range of time delays. 
 For example for $t_d = 2.9^{\pm0.4}$~Gyr the resulting range is $r_0= 2.04^{\pm0.08} \times 10^{11} $~cm.  
These initial separations are smaller than the main-sequence radii of main sequence progenitors of the neutron-stars  which are typically $>3 \times 10^{11}$~cm. This implies that the binaries have undergone a common envelope phase. 
This narrow range of initial separations seems to suggest that  this 
 common envelope phase ends with a  very narrow range of separations  and that somehow the formation of the second neutron star, by a supernova explosion of  or via another mechanism \citep{PiranShaviv05} does not disrupt this significantly. 

It is important to recall that the last conclusion that follows from the result concerning the narrow distribution of time delays  depends critically on the lack of non-Collapsar sGRBs at $z>1.2$, a distant at which \swift becomes insensitive to the lower
end of the sGRB luminosity distribution. A somewhat wider time delay distribute would have arisen if $z>1.2$ non-Collapsar sGRBs would have been observed by \swiftns. It is possible that such bursts have been detected but their redshift was not measured and hence they were not included in the sample. This would have changed significantly the conclusion concerning the narrow range of initial separation.

As ns$^2$ mergers (or bhns mergers), which are the most likely sources of sGRBs,  are the prime targets  of GW detectors
we turn to discuss the implications of our results  to GW detection. The approximate 
detection horizon\footnote{We assume here that the horizon is spherical. In reality the GW detection horizon is larger
along the rotation axis of the system and if, as expected, GRBs are along these direction then a chances for a coincident detection of a GW single and a GRB are larger. Of course independent of that the detection of a GRB increases significance of a GW detection at the same position and time and as such the corresponding detection horizon might be even further. } of ns$^2$ merger by Advanced LIGO/Virgo and other planned advanced GW detectors is
$300$ Mpc.  As the estimated local rate depends on the lower limit for the luminosity taken { we must refer to a minimum luminosity value when predicting detection rates.}
With our ``canonical" estimate of the lowest peak luminosity $5 \times 10^{49}$~erg/s
we expect within this  radius $0.06\pm0.03 yr^{-1}$ 
sGRBs detected by \swift and $0.4\pm0.2 yr^{-1}$ detected by Fermi/GBM. 
(see table \ref{tbl:GW_rates} for the rates for two other lower limits on  peak luminosity: the lowest observed peak luminosity by \swift, $ 2.2 \times 10^{50}$~erg/s, and an optimistic lower limit of $10^{49}$~erg/s.). 
These values reflect probabilities for simultaneous detection of a sGRB with a GW signal \citep{KP93}.
The estimated GW detection horizon for a bhns merger is $\sim 1$ Gpc. { The expected joint detection rate is $0.7\pm0.3 yr^{-1}$ and $5\pm2 yr^{-1}$ for \swift and Fermi respectively for the canonical lowest peak luminosity $5 \times 10^{49}$~erg/s (see table \ref{tbl:GW_rates}).}

When considering the overall detection rate of the GW detectors one has to consider the fact that sGRBs are most likely beamed. Using  a typical beaming factor of $f_b^{-1} \approx 30$ \citep[see also][]{Fong2012} and the ``canonical" lower limit on the peak luminosity of $5 \times 10^{49}$~ erg/s the event rate within a radius of $300 Mpc$ becomes $14_{-7}^{+8} yr^{-1}$ and the corresponding rate of detection of bhns mergers with a GW detection horizon of 1 Gpc is $5.1_{-2.4}^{+3.0}\times 10^2$ (see table \ref{tbl:GW_rates} for other values of the lowest peak luminosity).

\begin{table}
\begin{tabular}{|ll|lll|}
\hline
horizon & $L_{min}$ 		& GW det. alone		& Fermi 		& \swift  \\
$[Gpc]$ 	& $[$erg/s $]$		& $ yr^{-1}$  		& $ yr^{-1}$ 		& $ yr^{-1}$  \\
\hline
0.3 	& $10^{49}$ 		& $64_{-36}^{+53}$	& $1.0_{-0.5}^{+0.7}$ 	& $0.14_{-0.07}^{+0.11}$ \\
	& $5 \times 10^{49}$ 	& $14_{-7}^{+8}$ 	& $0.4\pm0.2$ 		& $0.06\pm0.03$ \\
	& $2.2 \times 10^{50}$ 	& $3.7\pm1.4$		& $0.11\pm0.04$ 	& $0.02\pm0.01$ \\
1 	& $10^{49}$ 		& $2.4_{-1.3}^{+1.9}\times 10^3$ & $6_{-3}^{+4}$  	& $0.8_{-0.4}^{+0.6}$ \\ 
	& $5 \times 10^{49}$ 	& $5.1_{-2.4}^{+3.0}\times 10^2$ & $5\pm2$ 		& $0.7\pm0.3$ \\
	& $2.2 \times 10^{50}$ 	& $138\pm53$		& $5_{-2}^{+1}$ 	& $0.5_{-0.2}^{+0.1}$ \\
\hline
\end{tabular}
 \caption{The expected GW detection rates with Advanced LIGO/Virgo for ns$^2$ mergers (detection horizon $\sim$ 0.3 Gpc) or bhns mergers (detection horizon $\sim$ 1 Gpc), for different lower limits on the peak luminosity $L_{min}$. Also given are the rates of a coincident detection of the GW and a detection of a sGRB by Fermi or by \swiftns. }
\label{tbl:GW_rates}
\end{table}

Finally it is interesting to consider the implications of these findings to the possibility that sGRBs are the sources of  heavy r-process material \cite{LattimerSchramm74,eichler1989,Frei99}.  This possibility received recently a lot of attention in view of the possible detection of a  IR signal 6 days (in the source rest frame) after the short GRB 130603B \citep{tanvir2013j,Berger+13}.
If the interpretation of this IR signal as macronova is correct it implies that $0.02 - 0.04$ r-process material were produced in this event \citep[][]{BarensKasen13,Hotokezaka+13,Grossman+14,Piran+14}. As there are about $10^{4} m_\odot$  of heavy ($A\>110 $) r-process material in the Milky Way, a few$ \times ~10^5$ such merger events are required to produce this material. Integration of the sGRB rate and using an effective density  of $0.01$ Mpc$^{-3}$ Milky Way like galaxies we obtain that there were $\approx 1.4 \times 10^4$ mergers beamed towards us within the Milky Way.  If all mergers produces a similar amount of r-process material than  a modest beaming factor of $f_b^{-1} \approx 20-30$ suffices to explain the observed abundance of these elements.  A critical consideration when proposing ns$^2$ mergers as sources of heavy r-process material is the observations of such elements in some very low metallicity ($\log_{10}[Fe/H]\approx 3$) stars 
\citep{woolf95,Shetrone96,Burris00,Cayrel01, Hill02}. 
The $Eu/Fe$ ratio varies strongly at this low metallicity, which agrees with the possibility that the heavy r-process elements are produced in rare events, but the question arises are there mergers so early on.  
A constant 
time delay of $\sim 2 $ Gyr is incompatible with such early merger events. However, the best fit power law time delay model (which was the less favored one in our analysis) yields $1000 (f_b^{-1}/30)$ mergers before redshift 5 and $10^4 (f_b^{-1}/30)$ mergers before redshift 3, and those might be sufficient to produce the required amounts of heavy r-process material sufficiently early on. 

\section*{Acknowledgments}
The research  was supported by an ERC advanced grant (GRBs) and by the  I-CORE 
Program of the Planning and Budgeting Committee and The Israel Science Foundation (grant No 1829/12).

\end{document}